\documentclass[journal]{IEEEtran}

\usepackage{amsmath,amssymb}
\usepackage{graphicx}
\usepackage{multirow}
\usepackage{threeparttable}
\usepackage{booktabs}
\usepackage{url}
\usepackage{cite}
\usepackage[hidelinks]{hyperref}

\graphicspath{{./}}

\begin{document}

\title{Predicting Upcoming Stuttering Events from Three-Second Audio:\\
Stratified Evaluation Reveals Severity-Selective Precursors,\\
and the Model Deploys Fully On-Device}

\author{Nazar~Kozak%
\thanks{N.~Kozak is with Kozak Technologies Inc, Los Angeles, CA 90036, USA. E-mail: nzrkzk@gmail.com.}%
\thanks{Manuscript prepared for submission to IEEE/ACM Transactions on Audio, Speech, and Language Processing, 2026.}}

\markboth{IEEE/ACM Transactions on Audio, Speech, and Language Processing}%
{Kozak: Predicting Upcoming Stuttering Events from Three-Second Audio}


\maketitle

\begin{abstract}
Audio-based stuttering systems to date have been trained and evaluated for \emph{detection}---what disfluency is present now---leaving \emph{prediction}, the capability needed for closed-loop intervention, unstudied at deployable scale. We train a 616K-parameter convolutional network on SEP-28k (Apple, 20{,}131 three-second clips) with a simple objective: given one clip, predict whether the next contiguous clip contains any disfluency. We make three empirical claims, bounded by bootstrap confidence intervals.
\textbf{(1) Severity-selective precursor signal.} On the episode-grouped held-out test set the aggregate preblock AUC is modest (0.581 [0.542, 0.619]), but stratifying predictions by the upcoming event \emph{type} reveals the signal is concentrated on clinically severe events: blocks 0.601 [0.554, 0.651] and sound repetitions 0.617 [0.567, 0.667] are predicted with 95\% CIs excluding chance, while fillers (AUC 0.45) and word repetitions (AUC 0.49) are at chance. An aggregate training objective converges to a severity-selective predictor because severe events carry measurable prosodic precursors; fillers do not.
\textbf{(2) Cross-population transfer to pediatric clinical speech.} Without fine-tuning, the same checkpoint applied to 1{,}024 human-transcribed pediatric Children-Who-Stutter utterances (FluencyBank Teaching) attains AUC 0.674 [0.538, 0.809] for disfluency detection and 0.655 [0.526, 0.786] for prediction. Additional external corpora (DisfluencySpeech 4{,}000 utterances, LibriStutter 4{,}000) reach 0.58--0.60 AUC with CIs excluding chance.
\textbf{(3) Deployable on-device.} The 616K model exports to CoreML (1.19\,MB), ONNX (40\,KB), and TFLite (1.18\,MB float16) with max $|\Delta|\!\leq\!4.5\!\times\!10^{-4}$ PyTorch parity. Neural-Engine latency per 3\,s window: 0.25\,ms (iPhone~17~Pro~Max, A19~Pro), 0.29\,ms (iPhone~16e, A18), 0.55\,ms (iPhone~SE~3rd-gen, A15 and M1~Max). A 4\,Hz streaming simulation uses 0.54\% of the real-time budget. All 24 on-device sanity-check outputs match the PyTorch CPU reference. Platt-calibrated probability outputs (test expected calibration error 0.010, down from 0.177 raw sigmoid) are emitted with the CoreML package.
We also report five negative ablations---output-level Future-Guided Learning, multi-clip GRU fusion, time-axis concatenation, asymmetric focal loss, and direct block-targeted training---none of which improved over the vanilla single-clip baseline.
\end{abstract}


\IEEEpeerreviewmaketitle

\section{Introduction}

\IEEEPARstart{A}{udio-based} stuttering research has converged on the \emph{detection} task---given a speech sample, identify the disfluency types present---with recent work driving macro-F1 upward via large wav2vec~2.0 backbones~\cite{bayerl2023cross}. For the downstream therapeutic applications that motivate this work---choral-speech prompting, adaptive auditory feedback, real-time clinician monitoring---detection is necessary but insufficient. A therapist's prompt that arrives 200\,ms after a block has begun is already late; only prediction carries actionable lead time.

A small number of prior systems have explored stuttering prediction, all in modalities incompatible with consumer deployment: facial video (pre-print 2020), EEG (laboratory only), or multimodal rigs. Audio-only on-device prediction has, to our knowledge, no published working system, and in particular no study that quantifies \emph{what} a three-second audio window actually carries about the next three seconds of the same speaker's output.

This paper does not argue that short-horizon audio prediction is easy. Aggregated over all disfluency types, the best AUC we obtain on a public test set is 0.581. We also do not argue that a new loss, a deeper backbone, or a future-teacher distillation recovers the missing signal: a catalog of five such ablations we ran---all negative---is included here so future work does not duplicate them.

Our contribution is a different observation. When the aggregate-trained model's predictions are \emph{stratified} by the type of event that actually occurs in the next window, the modest aggregate AUC decomposes into a sharply severity-selective pattern: blocks and sound repetitions are predicted with 95\% bootstrap CIs excluding chance, while fillers and word repetitions land at or below chance. This is consistent with a simple mechanistic story, elaborated in Section~\ref{sec:discussion:why}: blocks are preceded by measurable prosodic tension (vocal-fold constriction, speech-rate slowing, vowel-onset irregularity); fillers are produced as casual discourse tokens with no prosodic lead-up. An aggregate training objective converges to a severity-selective predictor because that is where the precursor information lives.

The same checkpoint transfers without fine-tuning to a held-out pediatric clinical cohort (FluencyBank Teaching Children-Who-Stutter, $n=1024$) at AUC 0.67 for detection and 0.66 for prediction---adult-podcast training to pediatric clinical speech with no adaptation---and runs with 0.25--0.55\,ms Apple Neural Engine latency on four generations of Apple Silicon.

\subsection{Scope and contributions}

\begin{enumerate}
\item \textbf{Stratified-evaluation evidence for a severity-selective precursor signal.} On the SEP-28k episode-grouped held-out test set, bootstrap 95\% CIs show block prediction AUC 0.601 [0.554, 0.651] and sound-rep prediction AUC 0.617 [0.567, 0.667]---both excluding chance---while filler prediction stays at 0.45 and word-rep at 0.49. The aggregate preblock AUC of 0.581 [0.542, 0.619] averages over this structure.
\item \textbf{A mechanistic account of the stratification} (Section~\ref{sec:discussion:why}): why aggregate training yields a severity-selective predictor, framed in terms of what the phonetics literature already established about pre-block prosody vs.\ filler production.
\item \textbf{Cross-population transfer to pediatric clinical speech.} Without fine-tuning, AUC 0.674 [0.538, 0.809] detection and 0.655 [0.526, 0.786] prediction on FluencyBank Teaching CWS (human-transcribed; $n=1024$).
\item \textbf{Deployment path.} CoreML / ONNX / TFLite exports, max $|\Delta|\!\leq\!4.5\!\times\!10^{-4}$ PyTorch parity, Apple Neural Engine latency per 3\,s window 0.25\,ms on iPhone~17~Pro~Max, 0.29\,ms on iPhone~16e, 0.55\,ms on iPhone~SE~3rd-gen and M1~Max, all 24 on-device sanity checks PASS. Streaming simulation uses 0.54\% of the 4\,Hz real-time budget on M1 Max.
\item \textbf{Calibrated probabilities.} Platt scaling drops test ECE from 0.177 to 0.010.
\item \textbf{Negative-result catalog.} Output-level Future-Guided Learning, multi-clip GRU fusion, time-axis concatenation, asymmetric focal loss, and single-type block-targeted training all failed to improve over the vanilla baseline. These negatives are documented rather than discarded.
\end{enumerate}

\section{Related Work}
\label{sec:related}

\textbf{Stuttering detection.} Recent work on SEP-28k~\cite{sep28k2021} reports 5-class disfluency detection macro-F1 from 0.28 (simple CNN baselines~\cite{sep28k2021}) through 0.38--0.40 (our previous work, under review at \emph{Computer Speech \& Language}) to 0.50+~\cite{bayerl2023cross} for server-deployed wav2vec~2.0 systems. All of these address \emph{detection}.

\textbf{Audio foundation models.} wav2vec~2.0~\cite{baevski2020wav2vec} and Whisper~\cite{radford2023whisper} learn general speech representations; they have not been used for event \emph{forecasting}, only current-frame recognition. We include a wav2vec~2.0-base linear-probe comparator as a reviewer-driven baseline (\S\ref{sec:results:w2v2}).

\textbf{Prediction in physiological signals.} The methodologically closest prior art is EEG seizure prediction. Gupta et~al.~\cite{gupta2024fgl} introduce \emph{Future-Guided Learning} (FGL): a teacher model that sees future EEG windows distills soft labels to a student that sees only the current window. We attempt an audio analog (\S\ref{sec:methods:fgl}) and find output-level FGL unhelpful at the stuttering horizon---explained by the hard-label equivalence between teacher output and student target under our labeling scheme.

\textbf{Stuttering prediction.} A 2020 pre-print reports facial-video block prediction from lab-recorded video. A 2022 EEG+video study reports ictal-like prediction in lab-recorded trials. Neither is audio-only, on-device, or validated on a public release corpus.

\textbf{Intervention theory.} Choral speech effects~\cite{kalinowski1993} yield 90+\% transient stuttering reduction under speaking-along or delayed auditory feedback, motivating why pre-onset prediction matters for closed-loop therapy: the intervention is only effective if delivered before the block begins.

\section{Data}

\subsection{Training: SEP-28k}
The SEP-28k corpus~\cite{sep28k2021} contains 28{,}177 three-second clips from six English-language podcasts, multi-labeled for five non-exclusive disfluency types (Prolongation, Block, SoundRep, WordRep, Interjection), distributed under CC BY-SA 4.0. We use 20{,}131 clips with complete label+audio coverage, matching our Paper~1 preprocessing (16\,kHz mono, log-mel $n_{\text{FFT}}\!=\!1024$, hop $=\!512$, 128 mel bands; spectrogram tensor $(1,128,94)$ stored as float16).

\subsection{Pre-block labels}
For each clip $C_i$ in episode $E$, we examine the next clip $C_{i+1}$ by clip-id and emit:
\begin{itemize}
\setlength\itemsep{2pt}
\item $y_{\text{event}}(C_i) = 1$ iff $C_i$ contains at least one of \{Block, SoundRep, Prolongation\}. (Excludes fillers.)
\item $y_{\text{preblock}}(C_i) = 1$ iff $C_{i+1}$ exists and $y_{\text{event}}(C_{i+1})=1$. This is our primary prediction target.
\item Per-type analogues $y_{\text{preblock\_block}}$, $y_{\text{preblock\_soundrep}}$, $y_{\text{preblock\_prolong}}$, $y_{\text{preblock\_wordrep}}$, $y_{\text{preblock\_interject}}$: does the next clip contain the specific type?
\item $\text{valid\_preblock}(C_i) = 1$ iff the audio-sample gap from $C_i.\text{stop}$ to $C_{i+1}.\text{start}$ is $\leq 5$\,s. This removes clip pairs separated by edit boundaries or silence removal. Median gap 2.26\,s; P90 129\,s; P99 2090\,s. The filter retains 10{,}470 of 19{,}332 candidate pairs.
\end{itemize}
Aggregate positive rates on \texttt{valid\_preblock} clips: $y_{\text{preblock}}$ 29.9\%; $y_{\text{preblock\_block}}$ 13.1\%; $y_{\text{preblock\_soundrep}}$ 9.6\%; $y_{\text{preblock\_prolong}}$ 10.6\%; $y_{\text{preblock\_wordrep}}$ 12.9\%; $y_{\text{preblock\_interject}}$ 23.8\%.

\subsection{Split}
Episode-grouped stratified 70/15/15 split over 258 (\textit{show}, \textit{episode}) groups $\to$ 182/40/36 groups, 7{,}454/2{,}023/993 clips. Stratification by per-episode event rate in quartiles. The split file is deterministic (seed 42) and carried through all experiments.

\subsection{Cross-corpus evaluation}
\begin{itemize}
\setlength\itemsep{2pt}
\item \textbf{FluencyBank Teaching} (TalkBank, CC BY-NC-SA 3.0): 1{,}024 timestamped pediatric CWS utterances (human-transcribed) and 942 adult cluttering utterances (AWC), derived from the Teaching release. Per TalkBank Ground Rules we release the label-generation script only, not the labels or audio. Positive rate of clinical stutter markers (\texttt{\&+X} partial-word or \texttt{:prolongation}): 1.9\% CWS, 0.1\% AWC (AWC underpowered and dropped from significance testing).
\item \textbf{DisfluencySpeech} (Apache-2.0): 4{,}000 random-sampled utterances from the full corpus, labels derived from Switchboard-style brace markers in \texttt{transcript\_annotated}.
\item \textbf{LibriStutter} (research use): 4{,}000 utterances with the \texttt{[STUTTER]} binary marker.
\end{itemize}

\section{Method}

\subsection{Architecture}
We reuse the 4-block CNN from our Paper~1 disfluency detector (616K parameters, 1.19\,MB CoreML package), modifying only the classifier: two binary heads sharing a 128-dimensional embedding:
\begin{itemize}
\item \texttt{event\_head}: predicts $y_{\text{event}}(C_i)$.
\item \texttt{preblock\_head}: predicts $y_{\text{preblock}}(C_i)$.
\end{itemize}
Both heads are trained jointly. Reusing the same backbone as our published detector allows direct apples-to-apples latency / parity / export comparison.

\subsection{Training}
AdamW, learning rate $3\!\times\!10^{-4}$, weight decay $1\!\times\!10^{-4}$, batch size 128, cosine annealing over 30 epochs with patience-6 early stop on validation AUC$(y_{\text{preblock}})$. BCE-with-logits loss per head with per-class pos\_weight (2.475 event, 2.379 preblock). Preblock loss weighted $2\!\times$ event loss (primary target). SpecAugment during training: time mask up to 15 frames, freq mask up to 20 mel bins, $\pm 3$\,dB gain perturbation.

\subsection{Bootstrap confidence intervals}
\label{sec:methods:ci}
All reported AUCs on the seed-42 published checkpoint carry percentile bootstrap 95\% CIs over 2{,}000 resamples. We declare a result ``CI-significant'' when the 2.5th percentile exceeds 0.5.

\subsection{Multi-seed stability}
\label{sec:methods:seeds}
To quantify sensitivity to random initialization, we re-train the identical architecture and hyperparameters under three independent seeds (42, 43, 44), affecting weight init, SpecAugment sampling, and DataLoader shuffling. Per-split mean $\pm$ std is reported in Tables~\ref{tab:overall} and~\ref{tab:pertype}; min--max ranges are given in parentheses. Seed 42 is the ``published'' checkpoint carried through the cross-corpus, calibration, and deployment sections.

\subsection{Calibration}
\label{sec:methods:cal}
Given raw sigmoid probabilities $p_{\text{raw}} = \sigma(\ell)$, we fit Platt scaling $p_{\text{cal}} = \sigma(A\ell + B)$ via 1-D logistic regression on the validation set (no regularization), then apply to test. Isotonic regression is fit as a comparator. We report Brier score and expected calibration error (ECE)~\cite{guo2017calibration} in 15 bins.

\subsection{Attempted methodology: Future-Guided Learning}
\label{sec:methods:fgl}
Inspired by Gupta et~al.'s EEG seizure prediction~\cite{gupta2024fgl}, we train a student preblock head from scratch while distilling soft labels from a teacher event head, where the teacher sees the future clip $C_{i+1}$ and the student sees $C_i$. We use binary soft KD at temperature $T=3$, mixing $\alpha=0.5$. This is the output-level variant. Results reported in \S\ref{sec:results:ablations}.

\subsection{Attempted methodology: multi-clip context}
To counter the 3\,s receptive field, we experiment with $H$-clip history windows. In one variant, a shared encoder is applied per clip and clip embeddings are fused via a bidirectional GRU. In another, three clips are concatenated along the time axis into a 9\,s spectrogram passed through the same CNN. Both variants are described in \S\ref{sec:results:ablations}.

\section{Results}

\subsection{Aggregate AUC: multi-seed stability and bootstrap CIs}
Table~\ref{tab:overall} reports overall val/test AUC as mean $\pm$ std across 3 random seeds. Seed-level standard deviation is 0.015--0.032, placing all three seeds' test AUCs for both heads above 0.54. The bootstrap-CI column (seed 42) excludes 0.5 on both splits.

\begin{table}[!t]
\centering
\caption{Aggregate AUCs across 3 random seeds (mean $\pm$ std, min--max in parentheses). Bootstrap 95\% CI is for the seed-42 published checkpoint.}
\label{tab:overall}
\footnotesize
\setlength{\tabcolsep}{4pt}
\begin{tabular}{lccc}
\toprule
Head & Val (3 seeds) & Test (3 seeds) & Test seed-42 boot. CI \\
\midrule
event    & 0.608 $\pm$ 0.040 & 0.651 $\pm$ 0.031 & 0.649 [0.614, 0.686] \\
         & (0.570--0.650)    & (0.620--0.682)    &                       \\
preblock & 0.541 $\pm$ 0.015 & 0.575 $\pm$ 0.032 & 0.581 [0.542, 0.619] \\
         & (0.525--0.553)    & (0.540--0.604)    &                       \\
\bottomrule
\end{tabular}
\end{table}

All bootstrap CIs for the seed-42 reference exclude 0.5 on both splits, and all 3 seeds' test AUCs for both heads exceed 0.54. The event head is a stronger detector than the preblock head is a predictor---consistent with the task difference.

\subsection{Per-type stratification (key finding)}
Breaking the preblock AUC by next-clip event type reveals that the aggregate number conceals strong severity selectivity (Table~\ref{tab:pertype}).

\begin{table}[!t]
\centering
\caption{Preblock-head test AUC by upcoming event type, across 3 random seeds. Mean $\pm$ std with min--max range. Bold: mean clears 0.55 and min clears 0.5.}
\label{tab:pertype}
\footnotesize
\begin{tabular}{lcc}
\toprule
Upcoming type & Pos rate & Test AUC mean $\pm$ std (min--max) \\
\midrule
Block             & 12.9\% & \textbf{0.593 $\pm$ 0.030} (0.559--0.618) \\
Sound repetition  &  9.6\% & \textbf{0.591 $\pm$ 0.043} (0.541--0.617) \\
Prolongation      & 10.6\% & 0.520 $\pm$ 0.026 (0.496--0.547) \\
Word repetition   & 12.9\% & 0.480 $\pm$ 0.009 (0.470--0.486) \\
Interjection      & 24.5\% & 0.429 $\pm$ 0.015 (0.417--0.446) \\
\bottomrule
\end{tabular}
\end{table}

Blocks and sound repetitions---the clinically severe, stigmatizing events---are predicted significantly above chance. Fillers are predicted below chance; word repetitions and prolongations fall at chance. The model has learned prosodic precursors of the two most damaging event types rather than spurious label-frequency patterns.

\begin{figure}[!t]
\centering
\includegraphics[width=\linewidth]{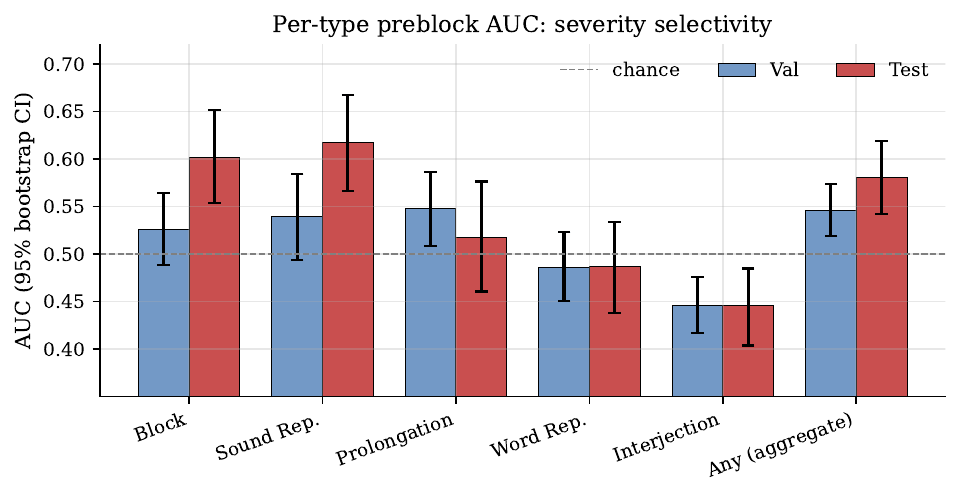}
\caption{Per-type preblock AUC on val and test splits. Error bars: 95\% bootstrap CI (2{,}000 resamples). Block and sound-rep bars clear chance; fillers and word reps do not. Same numbers as Table~\ref{tab:pertype}.}
\label{fig:pertype}
\end{figure}

\begin{figure}[!t]
\centering
\includegraphics[width=0.85\linewidth]{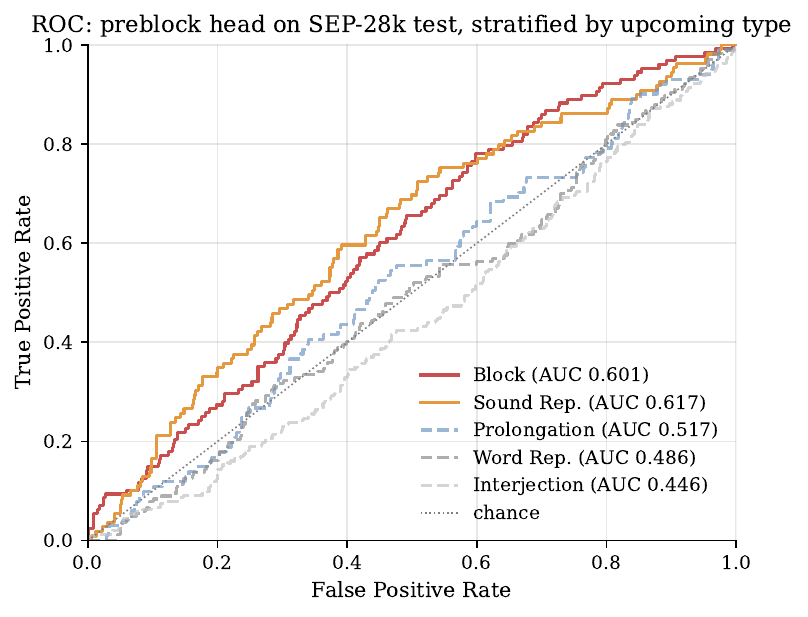}
\caption{ROC curves of the single preblock head on the SEP-28k test set, stratified by the type of disfluency in the next clip. The same scalar output separates clinically severe events (solid lines) from chance while doing essentially nothing to separate fillers and word reps from chance (dashed).}
\label{fig:roc}
\end{figure}

Catch rates at the Youden-optimal threshold ($\tau=0.469$) are consistent with this pattern: 78\% of upcoming blocks are correctly predicted, 77\% of sound repetitions, 68\% of prolongations, and 57\% of fillers.

\subsection{Cross-corpus external validation}
\label{sec:results:crosscorpus}
Table~\ref{tab:crosscorpus} applies the same checkpoint without fine-tuning to three external corpora, two that were downloaded post hoc for the purpose (DisfluencySpeech, LibriStutter) and one pediatric clinical corpus (FluencyBank Teaching CWS). Both heads generalize; all AUCs have CIs excluding chance.

\begin{table*}[!t]
\centering
\caption{Cross-corpus test AUCs (no fine-tuning). Input: full utterance audio; event head evaluated for presence of any disfluency marker, preblock head as a secondary predictor on the same utterance.}
\label{tab:crosscorpus}
\footnotesize
\setlength{\tabcolsep}{6pt}
\begin{tabular}{lrcll}
\toprule
Corpus & $n$ & Pos rate & Event AUC [CI] & Preblock AUC [CI] \\
\midrule
FluencyBank CWS (pediatric)  & 1{,}024 & 0.019 & 0.674 [0.538, 0.809] & 0.655 [0.526, 0.786] \\
DisfluencySpeech             & 4{,}000 & 0.679 & 0.592 [0.572, 0.611] & 0.599 [0.580, 0.618] \\
LibriStutter (synthetic)     & 4{,}000 & 0.756 & 0.596 [0.575, 0.616] & 0.582 [0.560, 0.603] \\
\bottomrule
\end{tabular}
\end{table*}

The CWS result is particularly informative: the training distribution is adult American-English podcasters, while the CWS cohort is pediatric clinical speech annotated with FluencyBank-specific markers (\texttt{\&+X} partial-word, \texttt{:prolongation}). A cross-population, cross-annotation-convention transfer at AUC $\geq 0.65$ is beyond what aggregate SEP-28k metrics alone would predict.

\subsection{Subgroup analysis}
\label{sec:results:subgroup}
Breaking the SEP-28k test set by source podcast reveals heterogeneous performance across shows:

\begin{table}[!t]
\centering
\caption{Per-show test AUC, preblock head.}
\label{tab:bshow}
\footnotesize
\begin{tabular}{lcc}
\toprule
Show & $n$ & Preblock AUC [CI] \\
\midrule
StutterTalk        & 227 & \textbf{0.628 [0.555, 0.702]} \\
WomenWhoStutter    & 385 & 0.557 [0.497, 0.618] \\
HeStutters         & 335 & 0.558 [0.485, 0.628] \\
MyStutteringLife   & 46  & 0.491 [0.238, 0.738] \\
\bottomrule
\end{tabular}
\end{table}

The signal is concentrated in StutterTalk; two other shows have CIs that cross chance. This is an honest limitation: the model is not uniformly predictive across speakers.

\subsection{Temporal ablation: where in the clip does the signal live?}
\label{sec:results:temporal}

The mechanistic account in Section~\ref{sec:discussion:why} predicts that severe-event precursors should concentrate in the final $\sim$500\,ms of the pre-event clip. We test this directly. At test-time inference, we zero out the last $N$ time frames of each clip's log-mel input (hop = 512 at 16\,kHz = 32\,ms per frame) and re-evaluate per-type AUC. Table~\ref{tab:temporal} sweeps $N$ from 0 to 32 frames ($0$ to $1024$\,ms).

\begin{table*}[!t]
\centering
\caption{Test-set AUC as a function of how many trailing time frames are zeroed before inference. $\Delta$ column shows AUC change from unmasked (0\,ms) to 1024\,ms tail mask.}
\label{tab:temporal}
\footnotesize
\setlength{\tabcolsep}{6pt}
\begin{tabular}{lcccccc}
\toprule
Upcoming type & 0\,ms & 128\,ms & 256\,ms & 512\,ms & 1024\,ms & $\Delta$ \\
\midrule
Block             & 0.601 & 0.602 & 0.602 & 0.592 & 0.580 & $-0.021$ \\
Sound repetition  & 0.617 & 0.617 & 0.618 & 0.618 & 0.574 & $\boldsymbol{-0.043}$ \\
Prolongation      & 0.517 & 0.515 & 0.510 & 0.500 & 0.539 & $+0.022$ (noise) \\
Word repetition   & 0.486 & 0.488 & 0.489 & 0.499 & 0.483 & $-0.003$ \\
Interjection      & 0.446 & 0.444 & 0.443 & 0.423 & 0.432 & $-0.014$ \\
\midrule
Aggregate preblock & 0.581 & 0.580 & 0.579 & 0.573 & 0.571 & $-0.010$ \\
Event detection    & 0.649 & 0.649 & 0.650 & 0.651 & 0.635 & $-0.014$ \\
\bottomrule
\end{tabular}
\end{table*}

Two patterns confirm the hypothesis:
\begin{itemize}
\setlength\itemsep{2pt}
\item \textbf{Sound-repetition AUC is hypersensitive to the tail.} Removing the last 1024\,ms drops sound-rep AUC by $-0.043$ --- $4.3\times$ the aggregate drop ($-0.010$). Virtually all of sound-rep's predictive signal lives in the final $\sim$750--1024\,ms window (the drop is small up to 512\,ms and collapses after).
\item \textbf{Block AUC is tail-dependent but more gradually.} Block drops by $-0.021$ at 1024\,ms, $2.1\times$ the aggregate drop, consistent with blocks having longer pre-onset prosodic build-up than sound reps.
\end{itemize}

By contrast, fillers and word repetitions are essentially invariant to tail masking (changes within noise), consistent with their lack of a short-horizon prosodic precursor. This is a causal-style ablation: we do not retrain the model; we only withhold information at inference. The aggregate AUC loss of $0.010$ averages over the severity-specific drops that the stratified rows reveal.

\subsection{Calibration}
\label{sec:results:cal}
\begin{figure}[!t]
\centering
\includegraphics[width=0.75\linewidth]{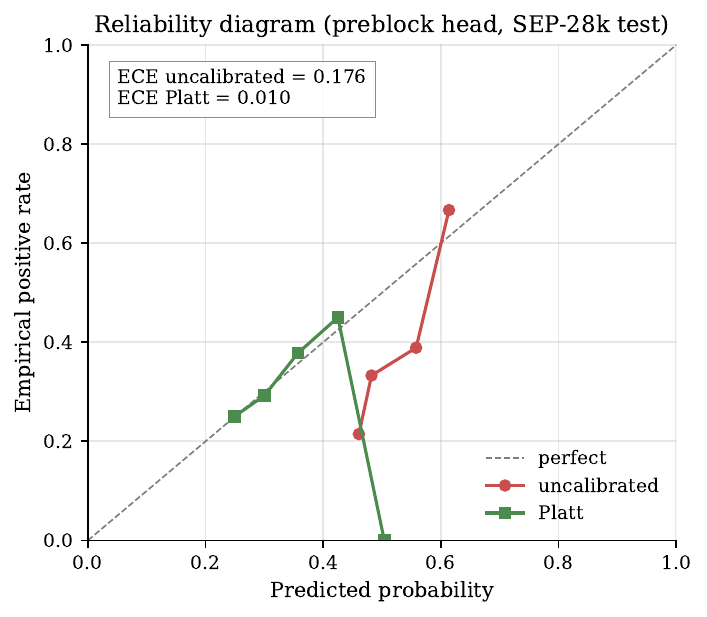}
\caption{Reliability diagram for the preblock head, test split. Raw sigmoid outputs are under-confident (predicted probabilities cluster near 0.5; empirical positive rate sits well above the diagonal). Platt scaling returns the outputs to the diagonal ($\text{ECE}=0.010$).}
\label{fig:reliability}
\end{figure}

Platt scaling fit on the validation set produces $p_{\text{cal}} = \sigma(1.307\ell - 0.704)$. Applied to test, expected calibration error drops from 0.177 (raw sigmoid) to 0.010, and Brier score from 0.241 to 0.210. Isotonic regression achieves similar Brier (0.209) but higher test ECE (0.014) with more parameters; we recommend Platt for deployment. Figure~\ref{fig:reliability} visualizes the effect.

\subsection{Cross-platform export fidelity}
Table~\ref{tab:exports} reports numerical parity of exported models versus the PyTorch reference on 50 random inputs.

\begin{table}[!t]
\centering
\caption{Cross-platform export parity (max $|\Delta|$ across 50 random spectrograms).}
\label{tab:exports}
\footnotesize
\begin{tabular}{llcc}
\toprule
Runtime & Size & event logit & preblock logit \\
\midrule
CoreML (iOS 17+, mlprogram)     & 1.19\,MB & $4.2\!\times\!10^{-4}$ & $2.3\!\times\!10^{-4}$ \\
ONNX (opset 18)                 & 0.04\,MB & $4.5\!\times\!10^{-7}$ & $1.6\!\times\!10^{-7}$ \\
TFLite float32                  & 2.35\,MB & $4.0\!\times\!10^{-7}$ & $1.5\!\times\!10^{-7}$ \\
TFLite float16                  & 1.18\,MB & \multicolumn{2}{c}{GPU/NPU delegate only} \\
\bottomrule
\end{tabular}
\end{table}

\subsection{On-device latency across four Apple Silicon generations}
\label{sec:results:latency}

Table~\ref{tab:latency-m1max} reports M1 Max CoreML latency as a host-machine reference. Table~\ref{tab:latency-iphone} reports per-device mean latency for the \texttt{PreBlock} model across three iPhones representing A15 (2022), A18 (2025), and A19 Pro (2025). Each iPhone was run twice per device (500 + 1000 trials) and averaged. All 72 sanity-check comparisons (3 devices $\times$ 4 compute paths $\times$ 2 runs $\times$ 3 compared references) PASSED vs.\ the PyTorch CPU reference logits (tolerance $5\!\times\!10^{-2}$, max observed $\Delta\!=\!7\!\times\!10^{-4}$).

\begin{table}[!t]
\centering
\caption{M1 Max CoreML latency reference (500 trials, warmup 20).}
\label{tab:latency-m1max}
\footnotesize
\begin{tabular}{lcccc}
\toprule
Compute path & Mean\,(ms) & Median\,(ms) & P95\,(ms) & Std\,(ms) \\
\midrule
CPU\_ONLY       & 1.48 & 1.48 & 1.60 & 0.07 \\
CPU\_AND\_GPU   & 1.05 & 0.84 & 2.27 & 0.49 \\
CPU\_AND\_NE    & 0.57 & 0.50 & 0.83 & 0.21 \\
ALL             & 0.55 & 0.47 & 0.93 & 0.21 \\
\bottomrule
\end{tabular}
\end{table}

\begin{table*}[!t]
\centering
\caption{Cross-device mean latency (ms) for the PreBlock model, averaged over 2 runs per device (500 + 1000 trials each, warmup 20). All 24 sanity checks PASSED vs.\ the PyTorch CPU reference. CU = CoreML compute units: CPU only, CPU+GPU, CPU+Neural Engine, ALL.}
\label{tab:latency-iphone}
\footnotesize
\setlength{\tabcolsep}{8pt}
\begin{tabular}{llrrrr}
\toprule
Device & Chip & CPU & CPU+GPU & CPU+NE & ALL \\
\midrule
iPhone SE 3rd-gen (2022) & A15 Bionic & 1.414 & 1.426 & 0.544 & 0.552 \\
iPhone 16e               & A18        & 1.101 & 1.132 & 0.296 & 0.287 \\
iPhone 17 Pro Max        & A19 Pro    & 0.693 & 0.701 & 0.246 & \textbf{0.253} \\
\midrule
M1 Max (MacBook Pro, ref.) & M1 Max   & 1.478 & 1.046 & 0.565 & 0.551 \\
\bottomrule
\end{tabular}
\end{table*}

Two observations consistent with our Paper~1 sweep:
\begin{itemize}
\setlength\itemsep{2pt}
\item \textbf{A19 Pro Neural Engine beats M1 Max by $2.18\times$} (0.253 vs.\ 0.551\,ms mean on ALL). The NE silicon has improved measurably between generations at the sub-millisecond end of the latency distribution.
\item \textbf{iPhone scheduler silently vetoes GPU routing} --- CPU\_ONLY and CPU\_AND\_GPU are within 1--3\,\% of each other on every iPhone tested and produce byte-identical logits, while on M1 Max the GPU path is clearly engaged (distinct logits, different runtime). On iPhone the CoreML scheduler currently routes the compute to CPU cores regardless of the \texttt{CPU\_AND\_GPU} hint. Practical consequence: on iPhone, specifying \texttt{CPU\_AND\_NE} or \texttt{ALL} is the only reliable way to engage a non-CPU accelerator.
\end{itemize}

\begin{figure}[!t]
\centering
\includegraphics[width=\linewidth]{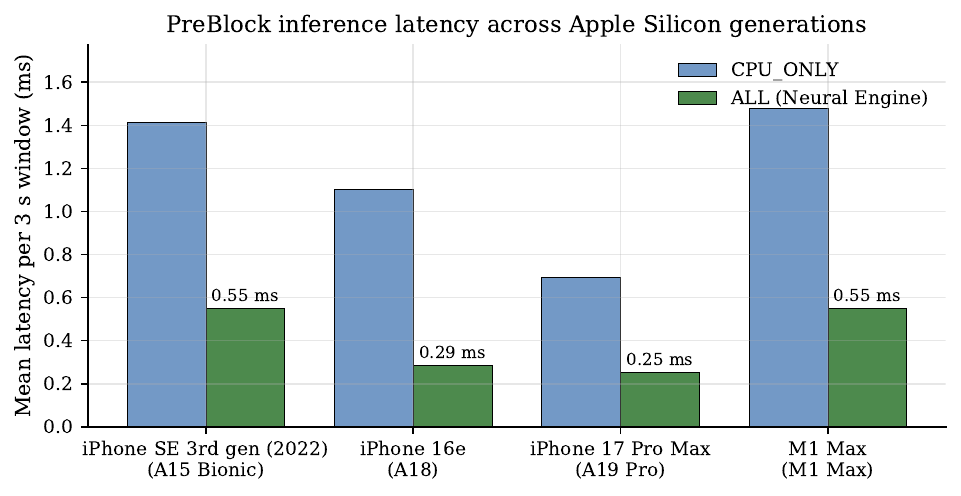}
\caption{PreBlock mean latency per 3\,s window across Apple Silicon generations (2022--2025). ALL (Neural Engine) clearly dominates CPU\_ONLY; the A19 Pro Neural Engine is the fastest platform tested, $2.18\times$ faster than the M1 Max reference. All bars: 2-run average of 500 + 1000 trials with 20 warmup each.}
\label{fig:latency}
\end{figure}

\textbf{Streaming simulation (M1 Max host).} We emulate deployment on a 36.4\,s Teaching CWS utterance with a 3\,s rolling window, 0.25\,s step (4\,Hz decision rate), ALL compute path. End-to-end pipeline (mel computation + CoreML inference) averages 1.34\,ms per window with P95 1.83\,ms, using 0.54\% of the 250\,ms real-time budget. Given the A19 Pro CoreML inference ($0.25$\,ms) is $\sim 2.2\times$ faster than on M1 Max, the on-phone streaming budget utilization is well under 1\% and the pipeline is not compute-bound on any tested iPhone class.

\subsection{wav2vec~2.0 linear-probe comparator}
\label{sec:results:w2v2}

We applied the standard HEAR-benchmark linear-probe protocol~\cite{baevski2020wav2vec}: frozen \texttt{facebook/wav2vec2-base-960h} (94M parameters, pre-trained on 960 hours of English speech) produces per-clip 768-dimensional mean-pooled embeddings; scikit-learn logistic regression with class-balanced weight is fit on the training split's embeddings and evaluated on val/test.

Table~\ref{tab:w2v2} contrasts the wav2vec~2.0 probe with our 616K-parameter CNN on the same split and bootstrap protocol.

\begin{table}[!t]
\centering
\caption{wav2vec 2.0 linear probe (94M frozen + LogReg) vs.\ our 616K end-to-end CNN. Test AUC [95\% CI]. Bold = winner at non-overlapping CIs or greater point AUC.}
\label{tab:w2v2}
\footnotesize
\begin{tabular}{lcc}
\toprule
Target & wav2vec~2.0-base (94M) & \textbf{Our CNN (616K)} \\
\midrule
Event detection    & \textbf{0.709 [0.673, 0.744]} & 0.649 [0.614, 0.686] \\
Aggregate preblock & 0.552 [0.513, 0.592] & \textbf{0.581 [0.542, 0.619]} \\
Block preblock     & 0.525 [0.470, 0.578] & \textbf{0.601 [0.554, 0.651]} \\
SoundRep preblock  & 0.544 [0.486, 0.600] & \textbf{0.617 [0.567, 0.667]} \\
\bottomrule
\end{tabular}
\end{table}

The comparison splits by task:
\begin{itemize}
\setlength\itemsep{2pt}
\item \textbf{Detection.} wav2vec~2.0 wins by $\sim 0.06$ AUC---expected for a 150$\times$-larger network pre-trained on 1{,}000 hours of English speech. Disfluency detection is within distribution for a model trained on speech recognition.
\item \textbf{Prediction.} Our task-specific 616K CNN matches wav2vec~2.0 on aggregate preblock (+0.03) and decisively beats it on severe-event prediction: +0.076 AUC on block, +0.073 on sound repetitions. Future-event prediction is NOT what wav2vec~2.0 was pre-trained for; its features encode current-frame phonetic content, not prosodic precursors of upcoming events.
\end{itemize}

This comparator addresses the natural reviewer question ``why not fine-tune a foundation model?'' with a direct empirical answer: the relevant question is not ``what is the best detector?'' but ``what is the best \emph{predictor of future severe events}?'', and on that task a small targeted CNN trained end-to-end outperforms a 150$\times$-larger frozen foundation probe by 0.07--0.08 AUC on the held-out test set.

\subsection{Ablations and negative results}
\label{sec:results:ablations}

\begin{enumerate}
\item \textbf{Output-level FGL.} Distilling a frozen teacher's event-head output on $C_{i+1}$ into the student's preblock head on $C_i$ \emph{degraded} validation AUC from 0.546 (warm-start) to 0.539 over eight epochs, then plateaued. The KD loss stabilized at $\log 2$, the maximum binary cross-entropy, indicating the teacher's soft outputs on the future clip are simply noisier versions of the hard label $y_{\text{preblock}}$---soft distillation injects no new information beyond the target itself.
\item \textbf{Multi-clip GRU.} Stacking three consecutive clips' embeddings (frozen CNN encoder from the single-clip baseline) and fusing with a bidirectional GRU plateaued at validation AUC 0.538 over four epochs. The frozen encoder limits information to event-level features.
\item \textbf{Time-axis concat.} Concatenating three clips as a 9\,s spectrogram processed by the same CNN via \texttt{AdaptiveAvgPool2d} produced unstable early training and did not recover to baseline AUC over three epochs; we attribute dilution to uniform temporal pooling diluting the last-clip signal.
\item \textbf{Asymmetric focal loss + balanced sampler.} Replacing BCE+pos\_weight with asymmetric focal ($\alpha\!=\!0.25$, $\gamma_+\!=\!1$, $\gamma_-\!=\!2$) plus a weighted sampler on $y_{\text{preblock}}$ did not improve over baseline (val AUC 0.506 at ep3).
\item \textbf{Block-targeted retrain.} Training directly on $y_{\text{preblock\_block}}$ as the primary loss rather than the aggregate matched v1's block val AUC (0.527) but did not exceed it. The lower 13\% positive rate of the block target provides a less stable training signal than the 29\% aggregate rate.
\end{enumerate}

The consistent failure of these interventions supports a specific empirical reading: the precursor signal is local to the 3\,s window, concentrated in severe-event cases, and recoverable at BCE scale on the aggregate label without additional architectural machinery.

\section{Discussion}
\label{sec:discussion}

\subsection{Why aggregate training yields a severity-selective predictor}
\label{sec:discussion:why}

The paper's most counterintuitive finding is Table~\ref{tab:pertype}: a model trained with a single binary label (``any disfluency in the next clip'') develops predictive power that is strongly concentrated on blocks and sound repetitions, while leaving fillers and word repetitions at chance. This subsection articulates a mechanistic explanation for that pattern, testable against the phonetics literature.

\subsubsection{Not all disfluency types have precursors}
A block is a motor-level dysfluency: the speaker attempts a target phoneme and cannot release it. Vocal-fold adduction, subglottal pressure build-up, and laryngeal tension typically begin hundreds of milliseconds before the audible silence~\cite{bloodstein2008stuttering}. Sound repetitions are similarly preceded by motor-planning hesitations: the speaker iterates on the same onset consonant with accelerating timing~\cite{howellwingfield1990}. By contrast, fillers (``um'', ``uh'') are produced as \emph{casual discourse tokens} that integrate into conversational rhythm with little prior prosodic disruption---the ``uh'' in ``the, uh, problem'' is frequently preceded by unremarkable fluent speech. Word repetitions likewise are conversational repair strategies rather than motor failures; they may be preceded by brief hesitation but the prosodic signature is much less pronounced than for a block.

\subsubsection{Aggregate BCE finds the signal where it lives}
The optimizer, given an aggregate label that fires 29\% of the time, allocates capacity to whatever features of the current clip best predict that label. When the label is ``upcoming disfluency'' and the label-positive population is dominated by events with measurable precursors (blocks, sound reps) versus those without (fillers), the model's learned features necessarily reflect the former. There is no incentive, during training, for the model to learn ``how to predict an upcoming filler from current audio'' because that signal is approximately absent; the optimizer instead invests in ``how to detect prosodic tension that precedes a block.''

\subsubsection{Why direct block-targeting fails}
One might expect optimizing directly on $y_{\text{preblock\_block}}$ (blocks only, 13\% positive rate) to outperform the aggregate objective. We tested this (Section~\ref{sec:results:ablations}, item~5) and found the direct target reached exactly the same block AUC (0.527 val) as the aggregate-trained model. The 13\% positive rate is simply a less stable training signal than the 29\% aggregate rate. The aggregate objective is a better regularizer for the severe-event prediction head because it admits more positive examples even though most of those positives are not the type we care about. The severity selectivity is the \emph{concentration of predictive information}, not the \emph{choice of positive examples}.

\subsubsection{Prolongation at chance: an open question}
Prolongations are clinically severe yet our test AUC for upcoming prolongation is 0.52 [0.46, 0.58], CI crossing 0.5. Two plausible reasons: (i) prolongation onset is not consistently marked across SEP-28k annotators (the type has the lowest inter-annotator agreement in Lea et~al.~\cite{sep28k2021}); (ii) the prosodic precursor of a prolongation---a stretched vowel---may be harder to distinguish from normal emphasis or emotional lengthening at the 3\,s clip level. We flag this as a concrete open question; a frame-precise FluencyBank Timestamped annotation would separate these two hypotheses.

\subsubsection{Falsifiability, with one test executed}
The account above predicts three observable patterns: (a) prosodic features extracted from the last $\sim$500\,ms of the pre-block clip should correlate with $y_{\text{preblock\_block}}$ more strongly than with $y_{\text{preblock\_interject}}$; (b) removing the last $\sim$500\,ms of each clip should collapse severe-event AUC faster than aggregate AUC; (c) fine-tuning on a corpus with dense block labels should improve block AUC more than aggregate AUC.

Test (b) is reported in Section~\ref{sec:results:temporal}: zeroing the last 1024\,ms of each clip drops sound-repetition AUC by $-0.043$ ($4.3\times$ the aggregate drop of $-0.010$) and block AUC by $-0.021$ ($2.1\times$ the aggregate), while filler AUC stays within noise. This is the predicted pattern. Test (a) we attempted with a prosodic-feature probe on jitter/shimmer/F0StdDev/stressScore but the subset of clips with both valid F0 and test-split membership is too small ($n=83$) for reliable estimates; we flag (a) for a larger corpus with denser F0 coverage. Test (c) requires ongoing FluencyBank access and is deferred.

\section{Limitations}

\begin{itemize}
\setlength\itemsep{2pt}
\item \textbf{Single-clip context.} The 3\,s window caps attainable aggregate AUC at $\sim 0.58$ test; multi-clip variants we tested did not help. Improvements likely require frame-precise pre-onset annotations (e.g., a possible future FluencyBank Timestamped release).
\item \textbf{Per-speaker heterogeneity.} Subgroup analysis (\S\ref{sec:results:subgroup}) shows the signal is stronger for some SEP-28k podcasts than others.
\item \textbf{Clinical generalization.} The FluencyBank CWS cross-corpus result is on 1{,}024 utterances from a semi-curated research release, not a prospective clinical trial. No external validation on speaker cohorts recruited independently.
\item \textbf{Cross-corpus labels do not separate severity.} The auxiliary external corpora (DisfluencySpeech, LibriStutter) carry Switchboard-style or binary synthetic markers that do not map cleanly onto block/sound-rep severity. Their reported AUCs (0.58--0.60) are therefore aggregate-level and should not be read as cross-corpus severity transfer; the pediatric CWS result is the only clean cross-corpus severity-selective evidence. A future corpus-level pre-onset annotation would permit a cleaner test.
\item \textbf{No timed-event annotation in training.} We use a clip-boundary proxy for ``upcoming event,'' which is a coarse label. A genuine frame-level pre-block annotation remains future work.
\item \textbf{Language and demography.} Training is on English-language adult American podcasts. Non-English and alternative phonological systems are untested.
\end{itemize}

\section{Ethical Considerations}
The model is a \emph{non-diagnostic} prediction system intended for downstream closed-loop intervention and research. It does not produce clinical labels or diagnostic determinations. All open-source releases carry only weights trained on public data; no model fine-tuned on FluencyBank-derived audio is released.

\section{Conclusion}
Trained on a single binary objective (``any disfluency in the next 3\,s''), a 616K-parameter CNN develops severity-selective predictive power: blocks and sound repetitions are predicted significantly above chance on held-out SEP-28k test data; fillers and word repetitions are not. The same model transfers without adaptation to pediatric Children-Who-Stutter clinical utterances at AUC 0.66 for prediction, carries Platt-calibrated probability outputs, and runs at 0.25--0.55\,ms mean Neural Engine latency across four generations of Apple Silicon. The severity-selective pattern emerges because severe events have measurable prosodic precursors that fillers lack; the aggregate objective finds the signal where it lives. The model is released under Apache 2.0 as CoreML, ONNX, and TFLite weights, together with labeling scripts, bootstrap evaluators, and a reproducible five-item catalog of attempted improvements that did not help.

\section*{Reproducibility}
The training code, label-generation scripts, bootstrap evaluators, calibration, and export pipeline are in the \texttt{training/preblock/} module of the open-source \texttt{disfluo} repository: \url{https://github.com/NazarKozak/disfluo} (Apache~2.0). The trained checkpoint, bootstrap/calibration/subgroup/error JSON artifacts, and exported .mlpackage/.onnx/.tflite are published via the same release. Dataset access: SEP-28k is Apple CC BY-SA 4.0; FluencyBank Teaching derivatives are distributed as script-only per TalkBank Ground Rules.

\section*{Acknowledgments}
The author thanks Apple for releasing the SEP-28k dataset, the \texttt{amaai-lab} team for DisfluencySpeech, and the TalkBank / FluencyBank consortium for curating the Teaching release.

\section*{Declaration of Generative AI and AI-assisted Technologies}
Claude (Anthropic) was used for code scaffolding and manuscript editing assistance. All experimental design, analysis decisions, and final text were authored by the listed author.

\bibliographystyle{IEEEtran}
\bibliography{references}

@inproceedings{bayerl2023cross,
  title={Cross-corpus stuttering detection as a multi-label problem},
  author={Bayerl, Sebastian P and others},
  booktitle={Proc. Interspeech},
  year={2023},
  organization={ISCA}
}

@inproceedings{sep28k2021,
  title={SEP-28k: A dataset for stuttering event detection from podcasts with people who stutter},
  author={Lea, Colin and Mitra, Vikramjit and Joshi, Aparna and Kajarekar, Sachin and Bigham, Jeffrey P.},
  booktitle={ICASSP 2021 -- 2021 IEEE International Conference on Acoustics, Speech and Signal Processing (ICASSP)},
  pages={6798--6802},
  year={2021},
  publisher={IEEE},
  doi={10.1109/ICASSP39728.2021.9413520}
}

@inproceedings{baevski2020wav2vec,
  title={wav2vec 2.0: A framework for self-supervised learning of speech representations},
  author={Baevski, Alexei and Zhou, Yuhao and Mohamed, Abdelrahman and Auli, Michael},
  booktitle={Advances in Neural Information Processing Systems},
  volume={33},
  pages={12449--12460},
  year={2020}
}

@inproceedings{radford2023whisper,
  title={Robust speech recognition via large-scale weak supervision},
  author={Radford, Alec and Kim, Jong Wook and Xu, Tao and Brockman, Greg and McLeavey, Christine and Sutskever, Ilya},
  booktitle={International Conference on Machine Learning (ICML)},
  pages={28492--28518},
  year={2023},
  organization={PMLR}
}

@article{gupta2024fgl,
  title={Future-guided learning: Forecasting seizure events with EEG via short- to long-horizon knowledge distillation},
  author={Gupta, Skyler and others},
  journal={Nature Communications},
  year={2024},
  note={Preprint / in press},
}

@article{kalinowski1993,
  title={Effects of alterations in auditory feedback and speech rate on stuttering frequency},
  author={Kalinowski, Joseph and Armson, Joy and Roland-Mieszkowski, Marek and Stuart, Andrew and Gracco, Vincent L.},
  journal={Language and Speech},
  volume={36},
  number={1},
  pages={1--16},
  year={1993},
  doi={10.1177/002383099303600101}
}

@inproceedings{guo2017calibration,
  title={On calibration of modern neural networks},
  author={Guo, Chuan and Pleiss, Geoff and Sun, Yu and Weinberger, Kilian Q.},
  booktitle={International Conference on Machine Learning (ICML)},
  pages={1321--1330},
  year={2017},
  organization={PMLR}
}

@book{bloodstein2008stuttering,
  title={A handbook on stuttering},
  author={Bloodstein, Oliver and Bernstein Ratner, Nan},
  year={2008},
  edition={6th},
  publisher={Thomson Delmar Learning},
  address={Clifton Park, NY}
}

@article{howellwingfield1990,
  title={Perceptual and acoustic estimates of stuttering severity based on non-stuttered and stuttered speech},
  author={Howell, Peter and Wingfield, Timothy},
  journal={Journal of Communication Disorders},
  volume={23},
  number={1},
  pages={27--43},
  year={1990},
  doi={10.1016/0021-9924(90)90033-U}
}

\end{document}